# Using Ordinal Data to Assess Distance Learning

Matthew Norris

mdnorris@gatech.edu

**1 INTRODUCTION**

By the end of March, 2020, every public school in the United States had shut its doors. Every instructor, from university professors to kindergarten teachers, had to now deliver educational content over the internet, and every student (and their family) needed to adapt to distance learning as the new status quo. Though K-12 education will eventually resume regular in-person learning, the point was made that schools must, at the least, plan for online education as an emergency necessity.

Academics in education and EdTech research have long recommended the flipped classroom model, which will be described shortly. Though not necessarily dependent on technology, formative assessment has also been largely hailed as a better alternative to traditional summative assessment. Despite slow adoption or even opposition, both methods would have worked together well during April and May of 2020. It also provided many students with experiences that, while not fully adopting formative assessment or flipped classrooms, contained some components of both. The purpose of the project was to create a survey that identified student attitudes about elements of flipped classrooms or formative assessments and determine if they had a statistically significant effect on outcomes. In actionable terms, it could isolate the tolerable or even enjoyable aspects of distance learning and bring those into the classroom, even if the classroom will never be flipped.

**2 RELATED WORK**

While there seems to be a consensus that flipped classrooms and formative assessment should be where education is headed, there is still research being done on both topics. As online learning is both experiencing growth in academic and business sectors, research to reduce the attrition rate and improve online education's capabilities are considered incredibly important. The concept of individual learning styles, while popular outside of those studying education, have been shown to have little to no effect on educational outcomes. A question referring to learning styles was included on the survey, and it will be briefly discussed to demonstrate the methods of analysis for this project.

**2.1 Flipped Classrooms**

A flipped or inverted classroom model can be simplistically defined as when "activities traditionally conducted in the classroom become home activities" (Akcayir, 2018), which is a very appropriate definition in the distance learning era. In more detail, the model asks students to engage with lectures and other educational media outside of the classroom, and time spent inside the classroom is utilized solving the roadblocks students have encountered during self-education. Research suggests that positive student interaction and reduced anxiety are demonstrated effects from implementing a flipped classroom (Akcayir, 2018).

Formative assessment is a crucial component of flipped classrooms' success in improving academic performance (Zainuddin, 2016). The two would likely have to be implemented concurrently, as research suggests that current testing methods would prevent flipped classroom adoption from being fully successful (Rotellar, 2016). Rotellar also discussed the high cost in time to faculty while they develop the new classroom environment, and that many students initially resist the transition. However, even the most driven of students (in this experiment, medical students), after participating in a flipped classroom environment, find that it is a richer environment for learning than the traditional classroom (Martinelli, 2017). Even in an elementary mathematics course, the flipped classroom has demonstrated it can encourage better outcomes (Lai, 2015).

The most recent papers emerging on the subject have been almost entirely qualitative in nature and use the required period of distance learning as their petri dish. One such paper found that online education in China was still able to follow a flipped classroom model and maintain student and teacher satisfaction (Yen, 2020). A paper from a Saudi Arabian university discusses how the training systems to implement flipped classrooms are not in place, but could provide a vital service during this time (Guraya, 2020).

**2.2 Formative Assessment**

One paper precisely defined formative assessment with its five primary goals:

> "1. Clarifying and sharing learning intentions and criteria for success;
>
> 2. Engineering effective classroom discussions and other learning task that elicit evidence of student understanding;
>
> 3. Providing feedback that moves learnings forward;
>
> 4. Activating students as instructional resources for one another; and
>
> 5. Activating students as the owners of their own learning." (Black, 2008)

Because it would replace summative assessment, which relies primarily on exams to determine a student's grade, there may be pushback with the belief that it allows students to take an easier path. However, it actually requires students to set their own standards of behavior and engage in self-monitoring behavior (Sadler, 1989) in addition to the five criteria above. Returning to the issue nearly a decade later, Sadler emphasized the feedback loop that is crucial for formative assessment to work (Sadler, 1999). As online learning becomes more common, developing learning communities with peers is difficult; performative assessment, as a process, facilitates interactions between geographically divided students (Gikandi, 2011).

The most recent research uses tech platforms like Moodle to help students in their self-assessments (McCallum, 2020). The students in McCallum's experiment reported that they felt more confident in their abilities to monitor and steer their own progress. It seems to be a valuable skill to build, especially during a time when the adults instructing them are not quite sure what might happen.

**2.3 Learning Styles**

The primary area of research that involves learning styles seems to be an attempt to build a platform or AI tutor that can tailor instruction to the student. In a paper that integrated formative assessment and learning style analysis, formative assessment was found to be effective, as was learning style (Wang, 2006). However, the only group that did succeed with their learning style were judged to be the reflective and observational learners; it seems quite possible that those students might have taken to self-assessment more quickly.

One paper discusses how self-reporting is no match for objective measurement: a study showed that self-reporting on what method is most comfortable had no correlation with methods that actually worked (Kirschner, 2016). Another study was unable to find any evidence of learning style efficacy, but was able to uncover study design errors in papers that found the theory effective (Pashler, 2009).

**2.4 Self-Motivation in Online Learning**

Research has been ongoing on the high attrition rates of online learning platforms. A primary factor is self-motivation, but motivation can be divided into extrinsic and intrinsic motivations. The paper suggested that external incentives to continue learning needed to be matched with a conscious effort to reduce student uncertainty and anxiety, which greatly affected intrinsic motivation (Chen, 2010). Another paper focused on the positive effects a competent instructor can have on an online learner's motivation, and the negative consequences that result from an imcompetent instructor (Selvi, 2010). Specific to mathematics instruction of children, research demonstrated the training in emotion regulation had positive educational effects (Cartwright, 2018).

**3 Methodology**

**3.1 Survey Design and Collection**

The questionnaire consisted primarily of 5-point Likert scale items. The first question asked in what year of high school the survey-taker was enrolled. The next thirteen questions were all of the Likert scale types. The Google Form had the numbers 1 through 5 selectable, and "Strongly Disagree" was associated with 1, while "Strongly Agree" was placed by the 5. Every Likert scale question began with the following request: "Please state to what degree you agree or disagree with the following statement." Many Likert scale questions reversed the order of a positive response to add to the validity of the design. The criticisms of that construction are primarily from confusing the survey-taker or exhausting their patience, which was unlikely given the short length and simplicity of the questionnaire.

The final four questions asked students to estimate their time spent studying and their math grades before and during distance learning. As was noted above, students do misreport, but the error that comes with self-reporting was likely not a weakness in this particular survey. Some schools stated they would not lower grades after beginning distance learning. Good work could improve grades, but falling behind would not harm them. The change in study hours and grades provided continuous variables to use with the ordinal data.

Each question was designed to be associated with a principal feature of formative assessment or flipped classrooms. The questionnaire's goal was to be short and easy to understand. Thus, each question aimed to be easy for the student while providing insight into how these concepts worked in a distance learning environment. The ideal survey results would provide individual responses useful for research and form useful groupings in the project's analysis phase. The full survey is available in the appendices

**3.2 Summary of Survey Results**

Seventy-one student responses came from the Kansas City, Denver, and Seattle metro areas. When the first completed surveys arrived, all public high schools had begun distance learning. Due to anonymity, there is no tally of the geographic distribution of responses, but there is an even distribution of grade levels. For the same reason, no efforts to ensure that the sample was representative of the United States population were feasible. Likely no rural students responded to the questionnaire. As the link's distribution came from one source in each metro area, it is possible respondents were more similar to one another than a random sample would have been.

**3.3 Method of Analysis**

The primary analysis focused on the distribution of the answers to each question and the statistical analysis and post-hoc tests performed on each question's responses. The results for each question will not be included for brevity's sake; only interesting or demonstrative analyses will be reported.

The Kruskal-Wallis test was performed on each question. Taking into consideration the ordinal (nonparametric) nature of the data and a smallish sample size, the Kruskal-Wallis test can be used to indicate whether the distribution of Likert responses is random or not. Pairwise post-hoc testing demonstrates the pairs of respondents that created a statistically significant result, and an effect size test can determine if an effect is low, moderate, or large.

To demonstrate the process, Question 9 on the survey is as follows:

"Please state to what degree you agree or disagree with the following statement: online learning is not my preferred learning style:" and asks for an ordinal response. By using the calculated difference between the pre-distance learning math grade and grade during distance learning as a dependent variable, I could perform the Kruskal-Wallis test. The results were:

*Kruskal-Wallis chi-squared = 6.755, df = 4, p-value = 0.1494*

As learning styles were criticized for not having an effect, it holds true in this experiment as well.

All of the other questions performed better, but some had other issues. Question 3 asked if the change in style of the math class made the student feel less confident in their ability to succeed in math.

*Kruskal-Wallis chi-squared = 9.0497, df = 4, p-value = 0.05987*

Due to uncertainty with sample size and a potential for an overly small or large group to throw off the test statistics, >90% was set as a threshold for rejecting the null hypothesis of the Kruskal-Wallis test. However, after running the Benjamini-Hochberg pairwise test on each group, the adjusted p-values went from significant to insignificant.

Finally, a "waffle plot" was made for each Likert item to see if the distribution might be grounds to dismiss a significant finding. Questions 10 and 14 were discarded because more than half the plot consisted of one Likert response. All other waffle plots will be included in the appendices. Here is Question 14:

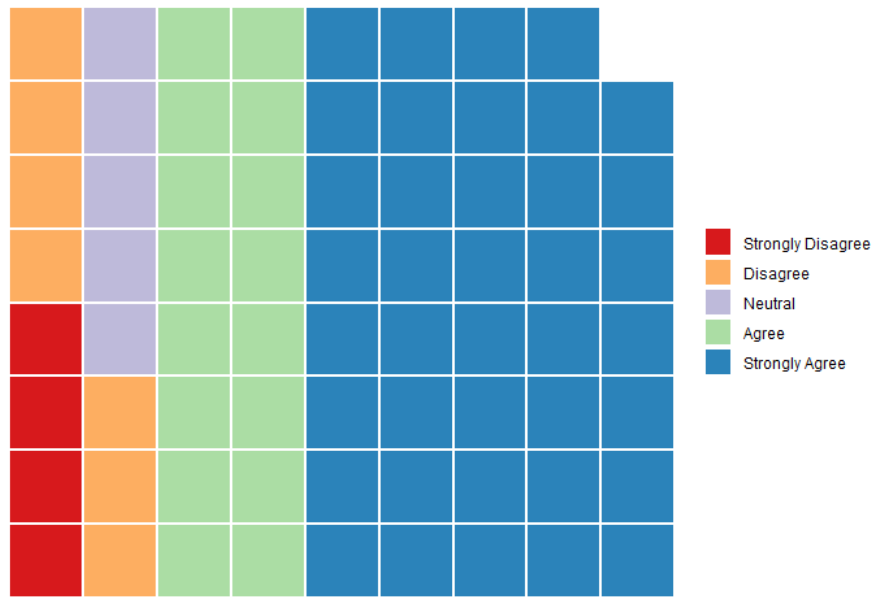

**4 Results**

The following six results were found to be statistically significant and relevant to the project's theme of looking for characteristics of formative assessment or flipped classrooms.

**4.1 Question 4 Analysis: Dependent Variable GPA Change**

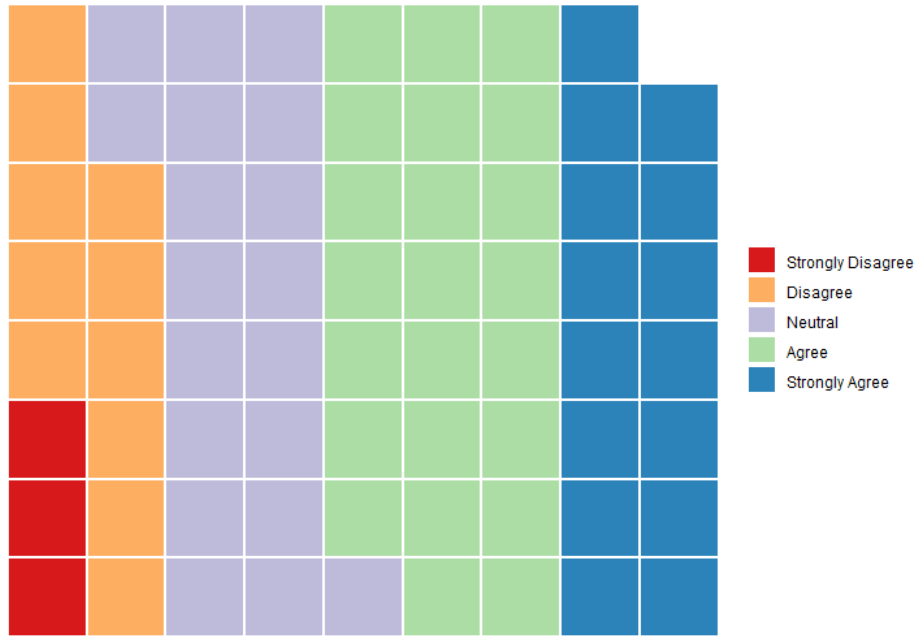

*Figure 1*—Waffle plot for Question 4

Though "Strongly Disagree" is weakly represented, the Kruskal-Wallis test is robust enough to handle that when the other four are more evenly distributed.

*The Kruskal-Wallis test chi-squared value was 20.844 with a p-value of 0.0003.*

*The Dunn Kruskal-Wallis pairwise test did not show any strange information.*

*The test for effect size returned an epsilon-squared statistic of .255, considered to be of large magnitude.*

## 4.2 Question 5 Analysis: Two Tests with GPA as Dependent Variable in the 1st and Study Time Difference in the Second

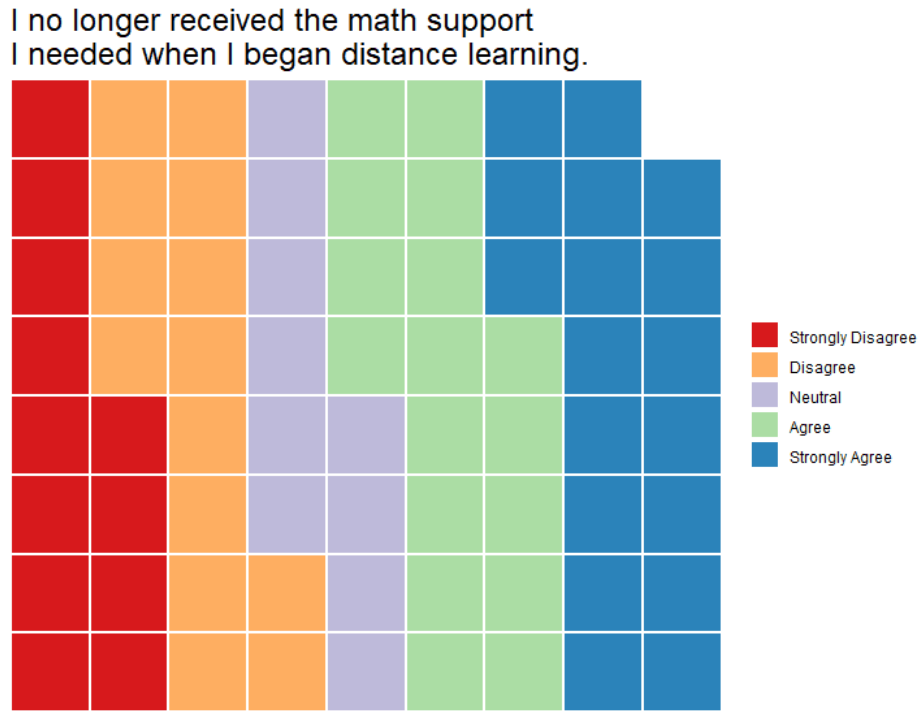

*Figure 2*—Question 4 Waffle Plot

For the GPA change:

*Kruskal-Wallis chi-squared = 17.769, df = 4, p-value = 0.001369*

*The Dunn Kruskal-Wallis pairwise test did not show any strange information.*

*The test for effect size returned an epsilon-squared statistic of .209, considered to be of large magnitude.*

For the study hours change:

*Kruskal-Wallis chi-squared = 21.288, df = 4, p-value = 0.0002776*

*The Dunn Kruskal-Wallis pairwise test did not show any strange information.*

*The test for effect size returned an epsilon-squared statistic of .262, considered to be of large magnitude.*

### 4.3 Question 6 Analysis: GPA as Dependent Variable

*Kruskal-Wallis chi-squared = 31.963, df = 4, p-value = 1.947e-06*

*The Dunn Kruskal-Wallis pairwise test did not show any strange information.*

*The test for effect size returned an epsilon-squared statistic of .424, considered to be of large magnitude.*

### 4.4 Question 7 Analysis

*Kruskal-Wallis chi-squared = 18.055, df = 4, p-value = 0.001204*

*The Dunn Kruskal-Wallis pairwise test did not show any strange information.*

*The test for effect size returned an epsilon-squared statistic of .213, considered to be of large magnitude.*

### 4.5 Question 12 Analysis with GPA as Dependent Variable

*Kruskal-Wallis chi-squared = 27.368, df = 4, p-value = 1.675e-05*

*The Dunn Kruskal-Wallis pairwise test did not show any strange information.*

*The test for effect size returned an epsilon-squared statistic of .354, considered to be of large magnitude.*

### 4.6 Question 13 Analysis with GPA as Dependent Variable

*Kruskal-Wallis chi-squared = 30.473, df = 4, p-value = 3.92e-06*

*The Dunn Kruskal-Wallis pairwise test did not show any strange information.*

*The test for effect size returned an epsilon-squared statistic of .354, considered to be of large magnitude.*

### 4.7 Summary of Findings

Questions 4, 5, 6, 7, 12, and 13 all were found to have statistical significance, or in Kruskal-Wallis terms, were found to have at least one group with stochastic dominance over the others. Listing out these six Likert items:

Q4: *my math teacher has done an excellent job developing an online curriculum for us.*

Q5: *I no longer receive the math support I need when I began distance learning.*

Q6: *to keep up in class, I feel the need to use other online learning tools to study.*

Q7: *I feel that the lack of peer interaction has caused my math performance to suffer.*

Q12: *online exams cause me more anxiety than in-person exams did.*

Q13: *my math teacher has made it clear that while exams are important, our completed coursework and involvement will do more for my final grade.*

Question 4 relates to both flipped classroom design and motivation.

Question 5 relates to feedback from the teacher, and also the potential feedback from peer interactions.

Question 6 could be indicative of a flipped classroom model, but the wording of keep up in class may have distorted it. However, half of all respondents selected "Strongly" disagree for that question.

For Question 7, over half of the respondents either selected "Agree" or "Strongly Agree". The wording of this question not only implies they miss their peers, but they are academically suffering from it.

Many people suffer from anxiety when taking exams, and Question 12 would likely be disagreed with if a flipped classroom model was in place. The responses, however, were evenly distributed.

Finally, Question 13's response would be very different if they were primarily graded on a summative assessment or formative assessment. Close to half "Agreed" or "Strongly Agreed".

To explore just a bit further, I performed a Polychoric Correlation test, which is suitable for ordinal variables, to judge the relationships between these variables. Question 13 is highly correlated with a correlation coefficient of .9 to Question 7, which implies that as responses move towards a classroom that is heavily based on formative assessment, so does the need to use online tools. This correlation only makes sense if there are some students engaging in the formative assessment model. The highest correlation on the table is between 5 and 7, which implies that as more students agree that their grades are suffering from a lack of peer interaction, more will also feel as though they lack support. This could be indicative of a flipped classroom model, where feedback from classmates is extremely valuable. The visualization of this Correlation Table and also a Factor Analysis and a Cluster diagram will be in the appendices.

**5 Limitations**

There were several limitations with this study. The sample size was on the border of being too small, and may have affected some of the statistical tests. Samples were pulled from three locations with a high chance that the survey was answered by someone similar to another person who responded, simply because people passed the questionnaire along to friends, who tend to be similar. The Kruskal-Wallis test is difficult to interpret beyond acknowledgment of difference, but the correlations did provide some suggestions for how they relate to each other. The visualizations in the appendix makes the existence of the relationship easier to acknowledge, but there is more that could be done. Finally, several fascinating questions about how the different grades differed in their responses were impossible due to the sample size.

**6 Future Work**

The question of "partially applied" flipped classroom and formative assessment methodologies is an interesting one, and the choices made to investigate it were sound. With a chance to send out a slightly modified survey, a larger sample size, some indicators that the sample was representative of the United States population, and some machine learning clustering algorithms, the analysis would likely return a rewarding answer.

**7 REFERENCES**

1. Akçayır, G., & Akçayır, M. (2018). The flipped classroom: A review of its advantages and challenges. Computers & Education, 126, 334-345.

2. Black, P., & Wiliam, D. (2009). Developing the theory of formative assessment. Educational Assessment, Evaluation and Accountability, 21(1), 5-31. doi:10.1007/s11092-008-9068-5

3. Cartwright, M., Kloos, H., Mano, Q., & Hord, C. (2018). Child guided math practice: The role of regulator emotional self-efficacy for children experience homelessness. CogSci.

4. Chen, A, & Jang, S. (2010) Motivation in online learning: testing a model of self-determination theory. Computers in Human Behavior, 26. 741-752.

5. Gikandi, J. W., Morrow, D., & Davis, N. E. (2011). Online formative assessment in higher education: A review of the literature. Computers & Education, 57(4), 2333-2351. doi:10.1016/j.compedu.2011.06.004

6. Guraya, S. (2020). Combating the COVID-19 outbreak with a technology-driven e-flipped classroom model of educational transformation. Journal of Taibah University Medical Sciences, 15(4), 253-254.

7. Kirschner, P. A. (2017). Stop propagating the learning styles myth. Computers & Education, 106, 166-171. doi:10.1016/j.compedu.2016.12.006

8. Lai, C.-L., & Hwang, G.-J. (2016). A self-regulated flipped classroom approach to improving students' learning performance in a mathematics course. Computers & Education, 100, 126-140.

9. McCallum, S. & Milner, M. (2020) The effectiveness of formative assessment: student views and staff reflections, Assessment & Evaluation in Higher Education, DOI: 10.1080/02602938.2020.1754761

10. Martinelli, S. M., Chen, F., DiLorenzo, A. N., Mayer, D. C., Fairbanks, S., Moran, K., . . . Schell, R. M. (2017). Results of a Flipped Classroom Teaching Approach in Anesthesiology Residents. J Grad Med Educ, 9(4).

11. Pashler, H., McDaniel, M., Rohrer, D., & Bjork, R. (2008). Learning Styles: Concepts and Evidence. Psychological Science in the Public Interest, 9.

12. Rotellar, C., & Cain, J. (2016). Research, Perspectives, and Recommendations on Implementing the Flipped Classroom. American Journal of Pharmaceutical Education, 80(2).

13. Sadler, D. (1998). Formative assessment: Revisiting the territory. Assessment in Education, 5(1), 77-85.

14. Sadler, D. (1989). Formative assessment and the design of instructional systems. Instructional Science, 18(2), 119-144.

15. Selvi, K. (2010). Motivating factors in online courses. Procedia Social and Behavioral Sciences, 2, 819-824.

16. Wang, K., Wang, T., W, W., & S, H. (2006). Learning styles and formative assessment strategy: enhancing student achievement in Web-based learning. Journal of computer assisted learning, 22(3), 207-217.

17. Yen, T. (2020). The Performance of Online Teaching for Flipped Classroom Based on COVID-19 Aspect. Asian Journal of Education and Social Studies, 8(3), 57-64.

18. Zainuddin, Z., & Halili, S. H. (2016). Flipped Classroom Research and Trends from Different Fields of Study. The International Review of Research in Open and Distributed Learning, 17(3). doi:10.19173/irrodl.v17i3.2274

# 8 Waffle, Factor Analysis, Cluster, and Polychoric Plots, and Survey

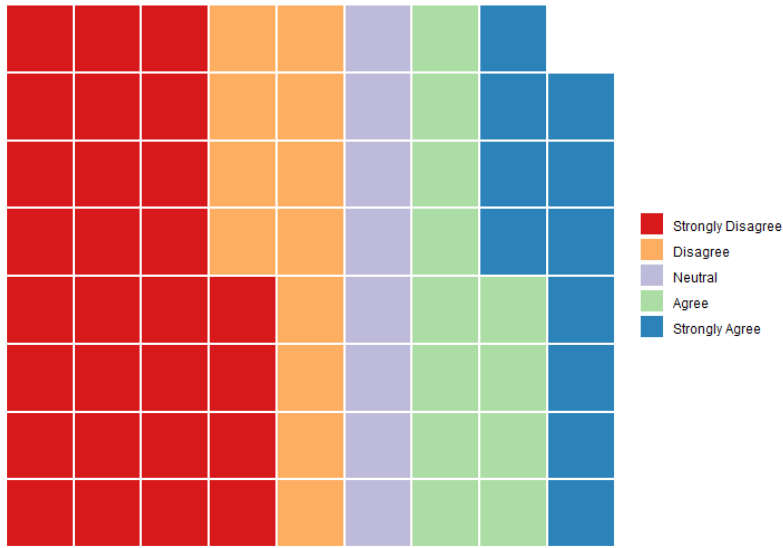

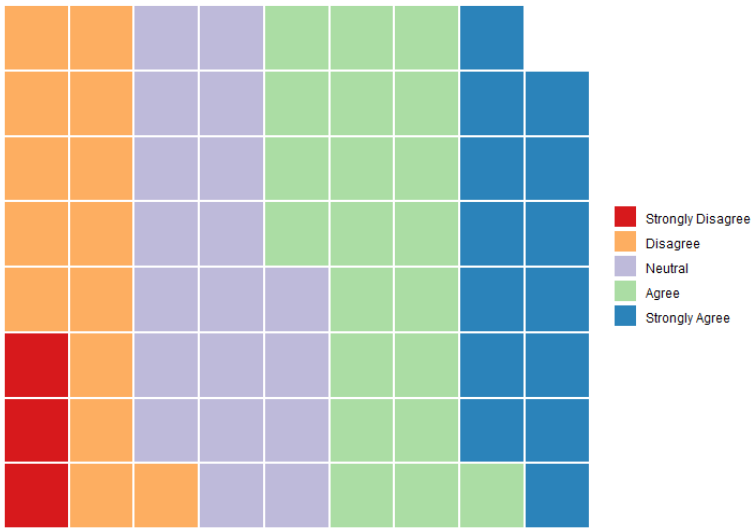

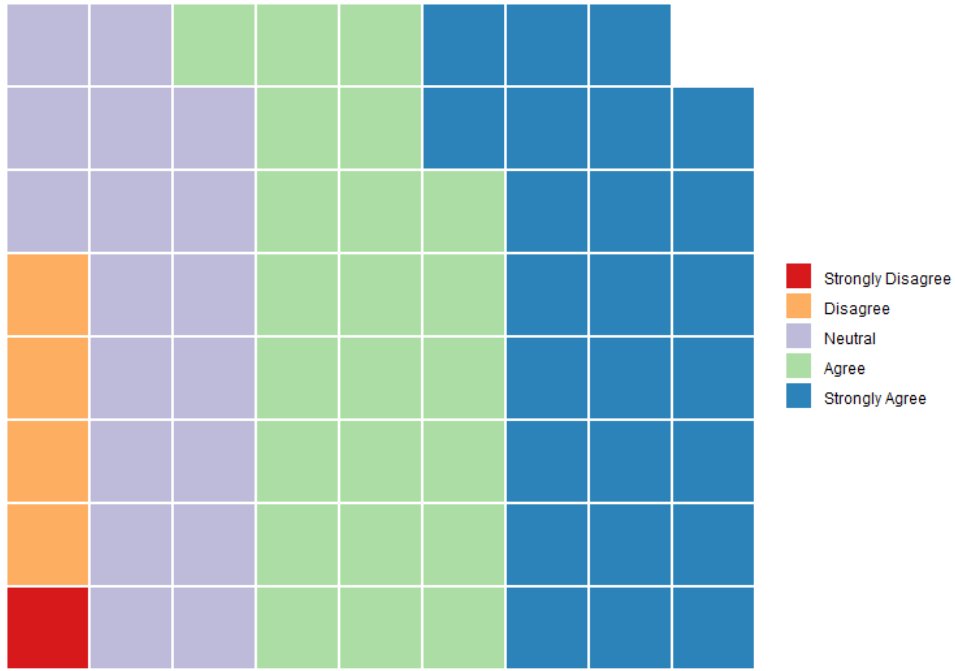

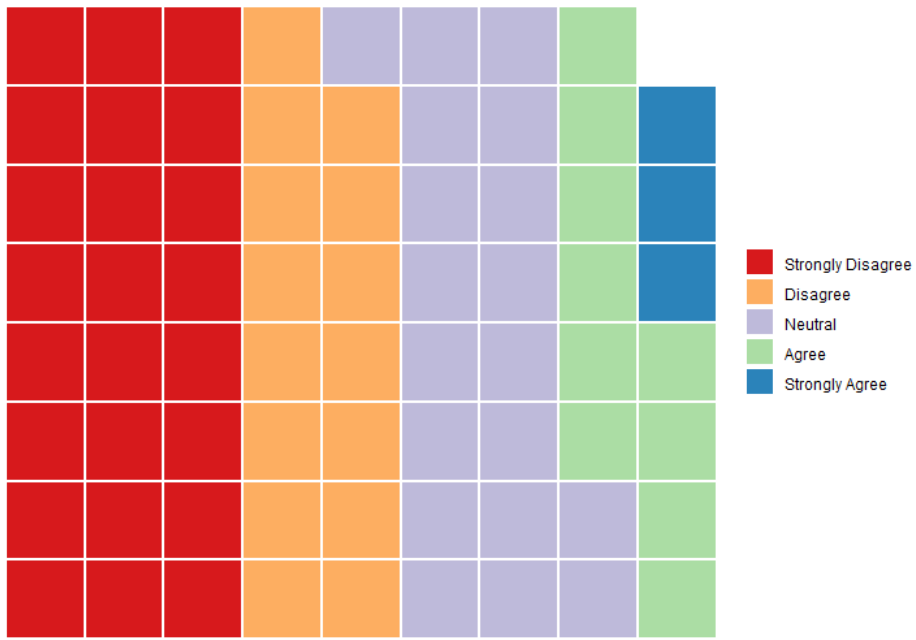

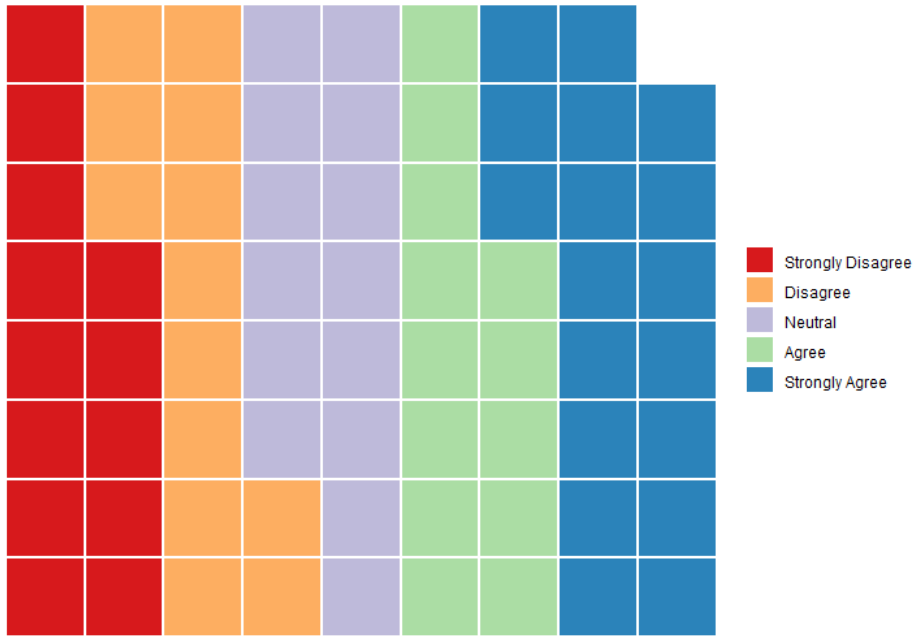

## 8.2 Correlation, Factor Analysis, and Cluster Plots

**Factor Analysis**

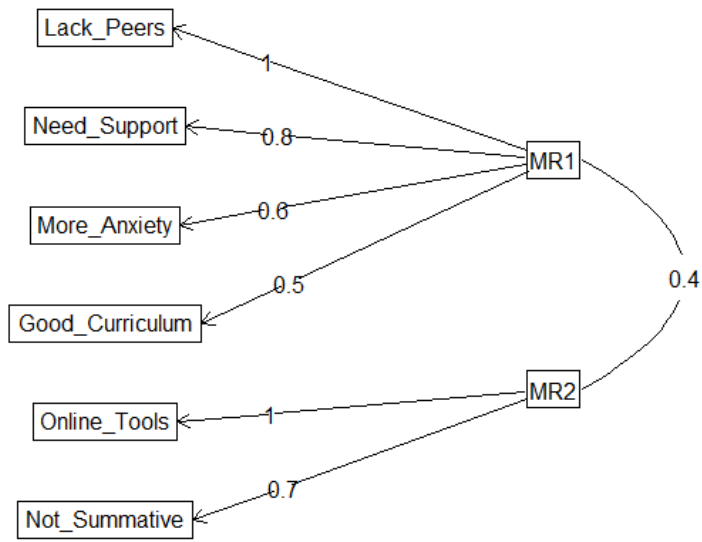

**ICLUST diagram**

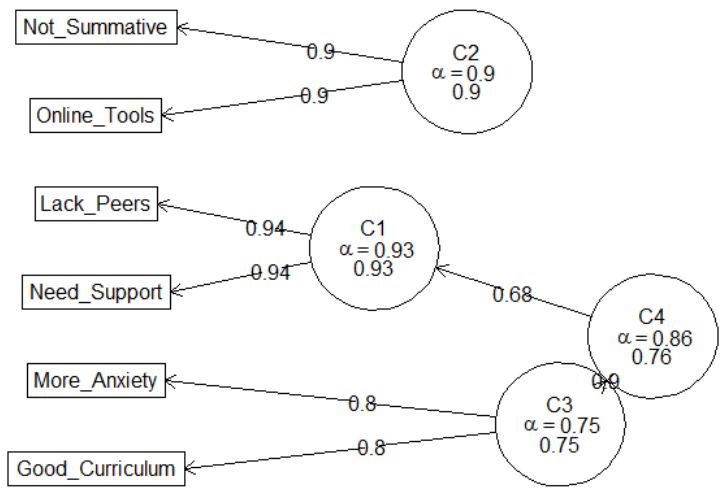

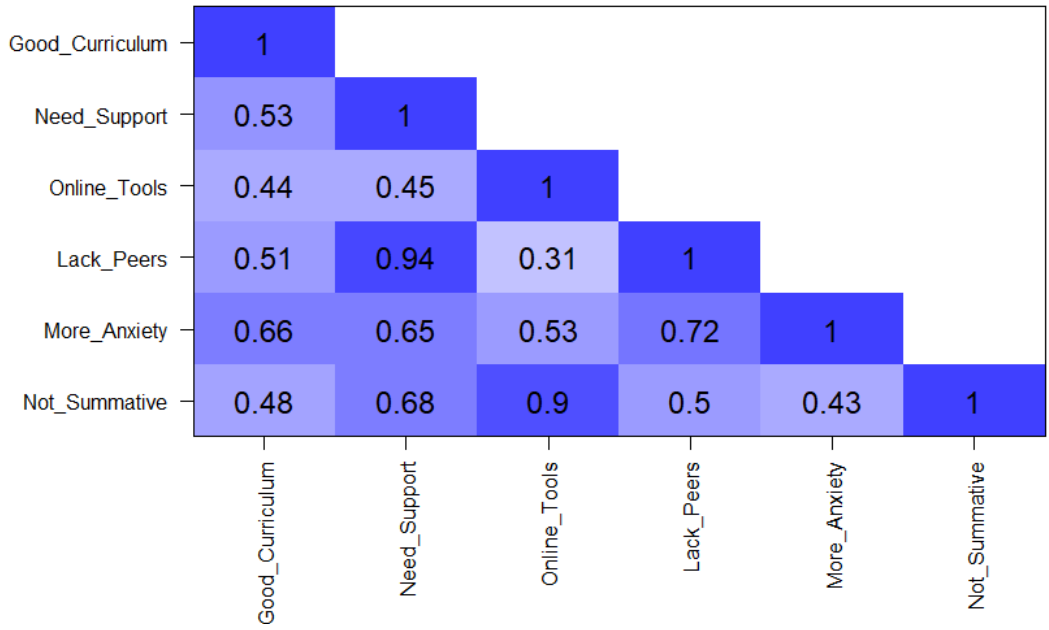

# Learning Methods Survey

This survey is completely confidential and will only be used to research online learning methods, math proficiency, and experiences with the new classroom structure.

* Required

**What grade are you currently in?** *

○ Freshman
○ Sophomore
○ Junior
○ Senior

Please state to what degree you agree or disagree with the following statement: I generally feel confident in my mathematical ability and can perform as well or better than my peers. *

|  | 1 | 2 | 3 | 4 | 5 |  |
|---|---|---|---|---|---|---|
| Strongly Disagree | ○ | ○ | ○ | ○ | ○ | Strongly Agree |

Please state to what degree you agree or disagree with the following statement: the change in style of my math class makes me feel less confident in my ability to succeed in math. *

|  | 1 | 2 | 3 | 4 | 5 |  |
|---|---|---|---|---|---|---|
| Strongly Disagree | ○ | ○ | ○ | ○ | ○ | Strongly Agree |

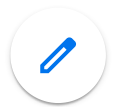

Please state to what degree you agree or disagree with the following statement: my math teacher has done an excellent job developing an online curriculum for us. *

| | 1 | 2 | 3 | 4 | 5 | |
|---|---|---|---|---|---|---|
| Strongly Disagree | ○ | ○ | ○ | ○ | ○ | Strongly Agree |

Please state to what degree you agree or disagree with the following statement: I no longer receive the math support I need when I began distance learning. *

| | 1 | 2 | 3 | 4 | 5 | |
|---|---|---|---|---|---|---|
| Strongly Disagree | ○ | ○ | ○ | ○ | ○ | Strongly Agree |

Please state to what degree you agree or disagree with the following statement: to keep up in class, I feel the need to use other online learning tools to study. *

| | 1 | 2 | 3 | 4 | 5 | |
|---|---|---|---|---|---|---|
| Strongly Disagree | ○ | ○ | ○ | ○ | ○ | Strongly Agree |

Please state to what degree you agree or disagree with the following statement: I feel that the lack of peer interaction has caused my math performance to suffer. *

| | 1 | 2 | 3 | 4 | 5 | |
|---|---|---|---|---|---|---|
| Strongly Disagree | ○ | ○ | ○ | ○ | ○ | Strongly Agree |

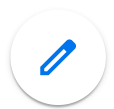

Please state to what degree you agree or disagree with the following statement: the online distance math curriculum is actually better than the in-class curriculum. *

| | 1 | 2 | 3 | 4 | 5 | |
|---|---|---|---|---|---|---|
| Strongly Disagree | ○ | ○ | ○ | ○ | ○ | Strongly Agree |

Please state to what degree you agree or disagree with the following statement: online learning is not my preferred learning style. *

| | 1 | 2 | 3 | 4 | 5 | |
|---|---|---|---|---|---|---|
| Strongly Disagree | ○ | ○ | ○ | ○ | ○ | Strongly Agree |

Please state to what degree you agree or disagree with the following statement: the flexibility of online instruction allows me to perform better in my math class. *

| | 1 | 2 | 3 | 4 | 5 | |
|---|---|---|---|---|---|---|
| Strongly Disagree | ○ | ○ | ○ | ○ | ○ | Strongly Agree |

Please state to what degree you agree or disagree with the following statement: I worry that online learning may cause me to drop behind where I should be in my math education. *

| | 1 | 2 | 3 | 4 | 5 | |
|---|---|---|---|---|---|---|
| Strongly Disagree | ○ | ○ | ○ | ○ | ○ | Strongly Agree |

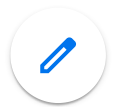

Please state to what degree you agree or disagree with the following statement: online exams cause me more anxiety than in-person exams did. *

|  | 1 | 2 | 3 | 4 | 5 |  |
|---|---|---|---|---|---|---|
| Strongly Disagree | ○ | ○ | ○ | ○ | ○ | Strongly Agree |

Please state to what degree you agree or disagree with the following statement: my math teacher has made it clear that while exams are important, our completed coursework and involvement will do more for my final grade. *

|  | 1 | 2 | 3 | 4 | 5 |  |
|---|---|---|---|---|---|---|
| Strongly Disagree | ○ | ○ | ○ | ○ | ○ | Strongly Agree |

Please state to what degree you agree or disagree with the following statement: it is more difficult for me to study before exams or start homework early now than it was during in-class instruction. *

|  | 1 | 2 | 3 | 4 | 5 |  |
|---|---|---|---|---|---|---|
| Strongly Disagree | ○ | ○ | ○ | ○ | ○ | Strongly Agree |

Before you began distance learning, estimate how many hours per week would you spend on math? *

Your answer

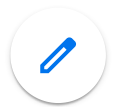

Since beginning distance learning, estimate how many hours per week do you spend on math? *

Your answer

In general, before distance learning, what would estimate your average math grade to be? *

○ A

○ A-

○ B+

○ B

○ B-

○ C+

○ C

○ C-

○ D

○ F

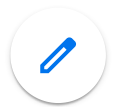

Not taking into account any special considerations your teacher may have announced that will help or prevent your grade from being hurt, what grade do you think your work deserves since beginning distance learning? *

- ○ A
- ○ A-
- ○ B+
- ○ B
- ○ B-
- ○ C+
- ○ C
- ○ C-
- ○ D
- ○ F

Page 1 of 1

Submit

Never submit passwords through Google Forms.

This form was created inside of Matthew Norris. Report Abuse

Google Forms

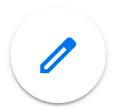